\documentclass{iopart}
\usepackage{iopams}
\begin{document}

\title{Generalized second-order partial derivatives of $1/r$}
\author{V Hnizdo}

\address{National Institute for Occupational Safety and Health,\\
Morgantown, West Virginia 26505, USA}
\ead{vhnizdo@cdc.gov}
\begin{abstract}
The generalized second-order partial derivatives of $1/r$, where $r$ is
the radial distance in 3D, are obtained using a result of the
potential theory of classical analysis. Some non-spherical-regularization
alternatives to the standard spherical-regularization expression for the derivatives
are derived. The utility of a spheroidal-regularization expression is illustrated
on an example from classical electrodynamics.
\end{abstract}
%\pacs{XXX}
%\maketitle
\section{Introduction}
The expression for the Laplacian  of $1/r$,
\begin{equation}
\nabla^2
\frac{1}{r} =-4\pi \delta(\bi{ r}),
\label{lap}
\end{equation}
where $r=|\bi{ r}|=(x_1^2+x_2^2+x_3^2)^{1/2}$ is the magnitude of a vector
$\bi{ r}=(x_1,x_2,x_3)$, the Laplacian  $\nabla^2$ is the differential operator $\partial^2/\partial x_1^2+\partial^2/\partial x_2^2+\partial^2/\partial x_3^2$ and $\delta(\bi{ r})
=\delta(x_1)\delta(x_2)\delta(x_3)$ is the 3D delta function,
is well-known and its correct use involves only the
elementary rules of the delta-function formalism.
The Laplacian (\ref{lap}) is useful in innumerable calculations of electrodynamics,
but sometimes the 2nd-order partial derivatives
$(\partial^2/\partial x_i\partial x_j)(1/r)$ themselves are needed.
Examples are the calculation of the electromagnetic fields of  point dipoles \cite{Griffiths} and solving  the Poisson equation for the difference between the Coulomb- and Lorenz-gauge vector potentials of a uniformly moving  point charge \cite{VH_EJP}.

While the expression for $(\partial^2/\partial x_i\partial x_j)(1/r)$ has been known for some time, its correct use is rather more intricate than that of $\nabla^2(1/r)$. This expression is usually written as \cite{Frahm}
\begin{equation}
\frac{\partial^2}{\partial x_i\partial x_j}\,\frac{1}{r} =
\frac{3x_ix_j-r^2\delta_{ij}}{r^5} -\frac{4\pi}{3}\,\delta(\bi{ r})\delta_{ij},
\label{frahm}
\end{equation}
which hides an important fact that an integration of the product of the first term on the right-hand side and a `well-behaved' test function over a 3-dimensional domain that includes the origin $\bi{ r}=0$ still has to be regularized because of the term's $1/r^3$ behaviour at the origin\footnote
{An expression for the (generalized) 2nd-order derivative of $1/r$ that takes this circumstance into account explicitly seems to have been  given first in \cite{Feri},
p 28.}.
The regularization assumed in (\ref{frahm}) is  of a specific, `spherical' kind, which can be effected in many equivalent forms, e.g.                              \begin{eqnarray}
\frac{3x_ix_j-r^2\delta_{ij}}{r^5}&\rightarrow&
{\rm w}\!\!\lim_{\epsilon\to 0}\frac{3x_ix_j-r^2\delta_{ij}}{(r^2+\epsilon^2)^{5/2}}
\label{reg_a}\\
&\rightarrow&{\rm w}\!\!\lim_{\epsilon\to 0}\frac{3x_ix_j-r^2\delta_{ij}}{r^5}\,
\Theta(r-\epsilon).
\label{reg_b}
\end{eqnarray}
Here, the symbol ${\rm w}\!\!\lim$ indicates the weak limit\footnote{
${\rm w}\!\!\lim_{a\to a_0} g_a(\bi{ r})=g(\bi{ r})$ iff $\lim_{a\to a_0}\int \rmd^3r\, f(\bi{ r}) g_a(\bi{ r})=\int \rmd^3r\, f(\bi{ r}) g(\bi{ r})$, where $g(\bi{ r})$ is in general a generalized function (distribution) and $f(\bi{ r})$ is any well-behaved test function.}
and $\Theta(\cdot)$ is the Heaviside step function. The form (\ref{reg_b})
is implemented automatically when spherical coordinates are used in  integration  and the angular integration is done first.

Relation (\ref{frahm}) was derived as a delta-function identity by Frahm \cite{Frahm}. In the present paper, we obtain it easily as a generalized (distributional)  derivative using a result of the potential theory of classical analysis. We also derive some `non-spherical' alternatives to (\ref{frahm}); while integration in spherical coordinates provides a straightforward implementation of the spherical regularization implied in (\ref{frahm}), the use of a non-spherical regularization may be more advantageous computationally in some applications.
An example from classical electrodynamics  where a spheroidal regularization is useful is given in Appendix.
Apart from the utility, awareness of non-spherical alternatives is important for the correct use of the standard, `spherical' expression.

Our treatment assumes no knowledge of the theory of generalized functions and generalized (distributional) derivatives beyond the elementary delta-function formalism, but it should help elucidate the operational meaning of some essentially non-classical mathematical objects\footnote
{For a general-function theoretic underpinning of our approach, the interested reader is directed to the canonical regularization method of Gelfand and Shilov  and the completeness theorem of generalized functions \cite{GS}.}. Presenting the 2nd-order partial derivatives of $1/r$ from the very beginning as generalized (distributional) derivatives should help avoid the pitfalls that would await anyone attempting to use
the expression (\ref{frahm}), or its very recent `generalization' \cite{Franklin} (remarked  on in the last section), in an integration in non-spherical coordinates.
This topic and its treatment are suitable for graduate and advanced undergraduate courses of electrodynamics.

A generalized-function treatment of the singularities
that may arise at the origin $r=0$ of a spherical (polar) coordinate system has been
presented recently in this Journal by Gsponer \cite{Gsponer}. In contrast to Gsponer,
our approach does not rely on any specific choice of coordinates.

\section{Derivation using derivatives of the potential of an extended density}
Let
\begin{equation}
\phi(\bi{ r})=\int \rmd^3r'\, \frac{\rho(\bi{ r}')}{R}, \qquad
R=|\bi{ r}-\bi{ r}'|
\label{phi}
\end{equation}
be the `potential' created by a density
$\rho(\bi{ r})$ that is assumed to be a `well-behaved' localized function of $\bi r$.
While the 2nd-order partial derivatives $\partial^2 \phi(\bi{ r})/\partial x_i\partial x_j$ perfectly exist at any point $\bi r$ if the  density $\rho(\bi{ r})$ is sufficiently `smooth'\footnote{
If the density $\rho_a(\bi{ r})$ has a sharp surface, the 2nd-order derivatives of the potential are discontinuous at that surface.},
they
cannot be calculated by a straightforward differentiating inside the integral that defines the potential because the resulting integrand is not integrable at the point
$\bi{ r}'=\bi{ r}$. A correct way of performing here the differentiation under the integral sign is given by the following formula:
\begin{eqnarray}
\fl \frac{\partial^2 \phi(\bi{ r})}{\partial x_i\partial x_j}
&=\lim_{\epsilon\to 0}
\int_{R>\epsilon}\rmd^3r'\, \rho(\bi{
r}')\frac{\partial^2}{\partial x_i \partial x_j}\,\frac{1}{R}-\frac{4\pi}{3}\,\rho(\bi{ r})\delta_{ij} \nonumber\\
\fl &=\lim_{\epsilon\to 0} \int \rmd^3r'\,
\rho(\bi{ r}')\frac{3(x_i-x_i')(x_j-x_j')-R^2\delta_{ij}}{R^5}\,\Theta(R-\epsilon)
-\frac{4\pi}{3}\,\rho(\bi{ r})\delta_{ij}.
\label{TS}
\end{eqnarray}
Here, the integration domain excludes a ball of radius $\epsilon$
and centre at $\bi{ r}'=\bi{ r}$; the 2nd line is a more explicit
transcription of the 1st line, with the integration domain $R=|\bi{
r}'-\bi{ r}|>\epsilon$ expressed by the Heaviside step-function
factor $\Theta(R-\epsilon)$ in the integrand.
An instructive derivation of formula (\ref{TS}) is given in the well-known text of Tikhonov and Samarskii \cite{TS} (they consider the most involved case $i=j$, but the generalization to any $i,j=1,2,3$ is straightforward).
This derivation uses the divergence theorem with a function that has only a $1/R$
singularity at the point $R=0$ and thus, unlike the informal proof in \cite{Frahm},
is fully legitimate in classical analysis\footnote{
As the informal proof in \cite{Frahm}, informal derivations of the full Laplacian relation (\ref{lap}) typically use
the divergence theorem with a function that has a $1/r^2$ singularity
at $r=0$; for a more rigorous alternative see \cite{VH00}.}.

Note that the regularization of the integral on the right-hand side of (\ref{TS}) is explicitly of the spherical kind. If the excluded
integration domain was assumed to have a  non-spherical shape,
the ensuing regularization would in general yield a different value for the integral, resulting in an incorrect value for the derivative in question.
The reason why different regularizations yield in general different values of the
regularized integral is that, because of the factor
$(\partial^2/\partial x_i\partial x_j)(1/R)$, the integrand goes through
large positive and negative variations near the point $R=0$. The integral thus can be made convergent only conditionally \cite{TS}.

Defining a generalized (distributional) derivative
\begin{equation}
\fl \frac{\bar{\partial}^2}{\partial x_i\partial x_j}
\,\frac{1}{R}\equiv {\rm w}\!\!\lim_{\epsilon\to
0}\frac{3(x_i-x'_i)(x_j-x'_j)-R^2\delta_{ij}}{R^5}\, \Theta(R-\epsilon)-\frac{4\pi}{3}\,\delta(\bi{ r}-\bi{ r}')\delta_{ij}
\label{genTSR}
\end{equation}
(denoted by a bar to distinguish it from a classical derivative \cite{Feri,Kanwal,Estrada}),
formula (\ref{TS}) can be written simply as
\begin{equation}
\frac{\partial^2 \phi(\bi{ r})}{\partial x_i\partial x_j}=
\int \rmd^3r' \rho(\bi{ r}')\frac{\bar{\partial}^2}{\partial x_i\partial x_j}
\,\frac{1}{R}.
\label{phigen}
\end{equation}
The generalized derivative (\ref{genTSR}) can thus  be seen as the mathematical operation using which the 2nd-order differentiation with respect to components of
$\bi r$ of a function defined as the integral with respect to $\bi{ r}'$ of an integrand involving the factor $1/R= 1/|\bi{ r}-\bi{ r}'|$ may be performed under the integral sign:
\begin{equation}
\frac{\partial^2}{\partial x_i\partial x_j}\int \rmd^3r'\,\frac{\rho(\bi{ r}')}{R}=\int \rmd^3r'\,\rho(\bi{ r}')\frac{\bar{\partial}^2}{\partial x_i\partial x_j}\,\frac{1}{R},
\end{equation}
where the `density' $\rho(\bi{ r})$ now plays the role of a well-behaved test function.

Expression (\ref{frahm}) is a special case $\bi{ r}'=0$ (so that $\bi{ R}\equiv \bi{ r}-\bi{ r}'=\bi{ r}$) of the generalized derivative (\ref{genTSR}),
\begin{equation}
\frac{\bar{\partial}^2}{\partial x_i\partial x_j}
\,\frac{1}{r}= {\rm w}\!\!\lim_{\epsilon\to
0}\frac{3x_ix_j-r^2\delta_{ij}}{r^5}\, \Theta(r-\epsilon)-\frac{4\pi}{3}\,\delta(\bi{ r})\delta_{ij}.
\label{genTSr}
\end{equation}
Adding expressions (\ref{genTSr}) with $i=j=1,2,3$, the non-delta-function terms cancel out, and we obtain the full Laplacian (\ref{lap}) as
\begin{equation}
\bar{\nabla}^2\frac{1}{r}=-4\pi\delta(\bi{ r}),
\end{equation}
where
\begin{equation}
\bar{\nabla}^2=
\frac{\bar{\partial}^2}{\partial x_1^2}
+\frac{\bar{\partial}^2}{\partial x_2^2}
+\frac{\bar{\partial}^2}{\partial x_3^2}
\end{equation}
is now the generalized Laplacian operator.

The generalized derivative (\ref{genTSr}) is  the weak limit $a\to 0$ of the corresponding classical derivative of the potential
\begin{equation}
\phi_a(\bi{ r})=\int \rmd^3r'\, \frac{\rho_a(\bi{ r}')}{R}, \qquad
R=|\bi{ r}-\bi{ r}'|
\label{phia}
\end{equation}
that is due to a localized density $\rho_a(\bi{ r})$ the
spatial extension of which depends on a parameter $a$ so that
\begin{equation}
\lim_{a\to 0}\int \rmd^3r\, f(\bi{ r})\rho_a(\bi{ r})= f(0),
\end{equation}
where $f(\bi{ r})$ is any well-behaved test function; this condition
is transcribed formally as
\begin{equation}
{\rm w}\!\!\lim_{a\to 0}\rho_a(\bi{ r})=\delta(\bi{ r}).
\end{equation}
This can be shown by calculating the limit $a\to 0$  of
the integral of the product of (\ref{TS}) with $\phi(\bi{ r})=\phi_a(\bi{ r})$
and a well-behaved function $f(\bi{ r})$. It suffices to consider the case $i=j=1$:
\begin{eqnarray}
\fl &\lim_{a\to 0}\int \rmd^3r\,f(\bi{ r})\frac{\partial^2
\phi_a(\bi{ r})}{\partial x_1^2}\nonumber \\
\fl &= \lim_{a\to 0}\int \rmd^3r f(\bi{
r})\lim_{\epsilon\to 0} \int \rmd^3r'\rho_a(\bi{
r}')\frac{3(x_1{-}x_1')^2{-}R^2}{R^5}
\Theta(R{-}\epsilon)-\frac{4\pi}{3}\lim_{a\to 0}\int
\rmd^3rf(\bi{ r})\rho_a(\bi{ r})\nonumber \\
\fl &= \lim_{a\to
0}\lim_{\epsilon\to 0} \int \rmd^3r\,f(\bi{ r})\int \rmd^3r'\, \rho_a(\bi{
r}')\frac{3(x_1-x_1')^2-R^2}{R^5}\,
\Theta(R-\epsilon)-\frac{4\pi}{3}\int \rmd^3r\,f(\bi{ r})\delta(\bi{ r})
\nonumber\\
\fl &=\lim_{a\to 0}\lim_{\epsilon\to 0}\int
\rmd^3r'\,\rho_a(\bi{ r}')\int \rmd^3r\,f(\bi{
r})\frac{3(x_1-x_1')^2-R^2}{R^5}\,
\Theta(R-\epsilon)-\frac{4\pi}{3}\,f(0)\nonumber\\
\fl &=\lim_{a\to 0}\int \rmd^3r'\,\rho_a(\bi{
r}')F(\bi{ r}')-\frac{4\pi}{3}\,f(0),
\label{TSf}
\end{eqnarray}
where
\begin{equation}
 F(\bi{ r}')=
\lim_{\epsilon\to 0}\int \rmd^3r\,f(\bi{ r})\frac{3(x_1-x_1')^2-R^2}{R^5}\,
\Theta(R-\epsilon).
\end{equation}
The function $F(\bi{ r}')$ is well-behaved, and thus
\begin{eqnarray}
\lim_{a\to 0}\int \rmd^3r'\,\rho_a(\bi{ r}')F(\bi{ r}') &=\int
\rmd^3r'\,\delta(\bi{ r}')F(\bi{ r}') \nonumber\\
&=F(0)\nonumber \\
&= \lim_{\epsilon\to 0} \int \rmd^3r\,f(\bi{ r})\frac{3x_1^2-r^2}{r^5}\,
\Theta(r-\epsilon).
\label{F}
\end{eqnarray}
The algebraic manipulations in (\ref{TSf}) are legitimate operations
of moving limits without changing their sequential order and of interchanging the orders of integration.

Equations similar to (\ref{TSf}) and (\ref{F}) obviously hold for all the  derivatives $\partial^2\phi_a(\bi{ r})/\partial x_i\partial x_j$, $i,j=1,2,3$. We thus have that, for any well-behaved test function $f(\bi{ r})$,
\begin{equation}
\fl \lim_{a\to 0}\int \rmd^3r\,f(\bi{
r})\frac{\partial^2 \phi_a(\bi{ r})}{\partial
x_i\partial x_j}=\lim_{\epsilon\to 0} \int \rmd^3r\,f(\bi{
r})\frac{3x_ix_j-r^2\delta_{ij}}{r^5}\, \Theta(r-\epsilon)-\frac{4\pi}{3}\,f(0)\delta_{ij},
\end{equation}
which is expressed formally as
\begin{eqnarray}
{\rm w}\!\!\lim_{a\to
0}\frac{\partial^2 \phi_a(\bi{ r})}{\partial
x_i\partial x_j}&=&{\rm w}\!\!\lim_{\epsilon\to
0}\frac{3x_ix_j-r^2\delta_{ij}}{r^5}\, \Theta(r-\epsilon)-\frac{4\pi}{3}\,\delta(\bi{ r})\delta_{ij}\nonumber\\
&=&\frac{\bar{\partial}^2}{\partial x_i\partial x_j}
\,\frac{1}{r}.
\label{frahm2}
\end{eqnarray}
Since the limit $a\to 0$ of the potential $\phi_a(\bi{ r})$ itself is the potential
$1/r$ of a point density $\delta(\bi{ r})$, this result is the formal
underpinning of a natural interpretation of the generalized derivative
 $(\bar{\partial}^2/\partial x_i\partial x_j)(1/r)$
as the 2nd-order derivative of the potential of a point source. Note that while the regularization used in (\ref{frahm2}) is of the spherical kind, the extended density $\rho_a(\bi{ r})$ that generates the potential $\phi_a(\bi{ r})$ does not have to have any particular symmetry as a function of $\bi{r}$. Using the powerful methods and results of the theory of generalized functions and derivatives, the result (\ref{frahm2}) can be obtained almost immediately (see \cite{VH04}, Appendix), but the approach adopted here required only a little more effort.

\section{Non-spherical regularizations}
\subsection{Spheroidal regularization}
Let us replace the spherical excluded integration domain used in (\ref{TS}), (\ref{genTSR}) and (\ref{genTSr}) by a spheroid
of semiaxes  $\epsilon/\gamma$, $\epsilon$ and $\epsilon$ along the  $x_1$, $x_2$ and $x_3$ axes, respectively, with the parameter $\gamma$ given by
\begin{equation}
\gamma=\frac{1}{\sqrt{1-v^2}}, \qquad  0<|v|<1.
\label{gamma}
\end{equation}
Regularization using such a domain may be suitable in applications involving
effects of special relativity, according to which a spherical charge of radius $\epsilon$
contracts to an oblate spheroid of this geometry when it is set in motion with a speed $v c$ along the $x_1$ axis. To find the modification of expression (\ref{genTSr}) for the generalized derivative $(\bar{\partial}^2/\partial x_i\partial x_j)(1/r)$ that such spheroidal regularization entails, we only need to evaluate the difference
\begin{eqnarray}
\fl
&\lim_{\epsilon\to 0}\int_{\gamma^2 x_1^2+x_2^2+x_3^2>\epsilon^2}\rmd^3r\,f(\bi{ r})\frac{3x_ix_j-
r^2\delta_{ij}}{r^5}
-\lim_{\epsilon\to 0}\int_{r>\epsilon}\rmd^3r\,f(\bi{ r})\frac{3x_ix_j-
r^2\delta_{ij}}{r^5}\nonumber\\
\fl &\quad =\lim_{\epsilon\to 0}\int_{{\cal U}_{\epsilon}}\rmd^3r\,f(\bi{ r})\frac{3x_ix_j-
r^2\delta_{ij}}{r^5},
\label{diff}
\end{eqnarray}
where
$f(\bi{ r})$ is again a well-behaved test function and the integration domain ${\cal U}_{\epsilon}$ is the region delimited by the oblate surface  $\gamma^2 x_1^2+x_2^2+x_3^2=\epsilon^2$ and the spherical surface $r^2=x_1^2+x_2^2+x_3^2=\epsilon^2$:
\begin{equation}
{\cal U}_{\epsilon}=\{(x_1,x_2,x_3);\gamma^2 x_1^2+x_2^2+x_3^2>\epsilon^2\cap x_1^2+x_2^2+x_3^2<\epsilon^2\}.
\end{equation}

As $\epsilon$ tends to zero, the integration domain ${\cal U}_{\epsilon}$
gets progressively smaller and closer to the origin $\bi{ r}=0$ so that, for any
$\bi{ r}\in {\cal U}_{\epsilon}$, $f(\bi{ r})\to f(0)$ as $\epsilon\to 0$. The right-hand side of (\ref{diff}) can therefore be written as
\begin{equation}
\lim_{\epsilon\to 0}\int_{{\cal U}_{\epsilon}}\rmd^3r\,f(\bi{ r})\frac{3x_ix_j-
r^2\delta_{ij}}{r^5}=f(0)\lim_{\epsilon\to 0}\int_{{\cal U}_{\epsilon}}\rmd^3r\,\frac{3x_ix_j-r^2\delta_{ij}}{r^5}.
\label{diffU}
\end{equation}
Here, the integral on the right-hand side can be evaluated easily in spherical coordinates. With $x_1$ axis as the polar axis and $\cos\theta=\xi$, we obtain
for $i=j=1$:
\begin{eqnarray}
\int_{{\cal U}_{\epsilon}}\rmd^3r\,\frac{3x_1^2-r^2}{r^5}
&=&2\pi\int_{-1}^1 \rmd\xi\,(3\xi^2-1)\int_{\epsilon/\sqrt{1+(\gamma^2-1)\xi^2}}^{\epsilon}
\frac{dr}{r}\nonumber \\
&=&\pi\int_{-1}^1\rmd\xi\,(3\xi^2-1)\ln[1+(\gamma^2-1)\xi^2]\nonumber\\
&=&2\pi\left(\frac{2}{v^2}-\frac{2\arcsin v}{\gamma v^3}-\frac{2}{3}\right).
\label{ij11}
\end{eqnarray}
The case $i=j=2$ gives
\begin{eqnarray}
\fl \int_{{\cal U}_{\epsilon}}\rmd^3r\,\frac{3x_2^2-r^2}{r^5}
&=\frac{1}{2}\int_{-1}^1\rmd\xi\int_0^{2\pi}\rmd\phi\,[3(1-\xi^2)\cos^2\phi-1]\ln[1+(\gamma^2-1)\xi^2]
\nonumber\\
\fl &=\frac{\pi}{2}\int_{-1}^1\rmd\xi\,(1-3\xi^2)\ln[1+(\gamma^2-1)\xi^2]\nonumber\\
\fl &=2\pi\left(\frac{1}{3}-\frac{1}{v^2}+\frac{\arcsin v}{\gamma v^3}\right),
\label{ij22}
\end{eqnarray}
and the same result obviously will be obtained for $i=j=3$.
The mixed cases $i\ne j$ will all yield zero on account of the integration with respect to the azimuthal angle $\phi$. A notable feature of the results (\ref{ij11}) and (\ref{ij22}) is that they are independent of $\epsilon$. Collecting all these results,
equation (\ref{diffU}) can be written as
\begin{equation}
\lim_{\epsilon\to 0}\int_{{\cal U}_{\epsilon}}\rmd^3r\,f(\bi{ r})\frac{3x_ix_j-
r^2\delta_{ij}}{r^5}=2\pi [g_{ij}(v)-\case{2}{3}\delta_{ij}]f(0),
\label{collect}
\end{equation}
where
\begin{equation}
g_{ij}(v)=\left\{
\begin{array}{ll}
2/v^2-(2/\gamma v^3)\arcsin v &  i=j=1\\
1 -1/v^2+(1/\gamma v^3)\arcsin v & i=j=2,3\\
0  &  i\ne j.
\end{array}
\right.
\end{equation}

This result establishes a generalized-function identity
\begin{equation}
\fl {\rm w}\!\!\lim_{\epsilon\to
0}\frac{3x_ix_j-r^2\delta_{ij}}{r^5}\,[\Theta(\gamma^2 x_1^2+x_2^2+x_3^2-\epsilon^2)
-\Theta(r-\epsilon)]=2\pi [g_{ij}(v)-\case{2}{3}\,\delta_{ij}]\delta(\bi{ r}),
\label{identity0}
\end{equation}
using which the difference (\ref{diff}) with any well-behaved function
$f(\bi{ r})$ can be evaluated immediately.
According to this identity
\begin{eqnarray}
\fl & {\rm w}\!\!\lim_{\epsilon\to 0}\frac{3x_ix_j-r^2\delta_{ij}}{r^5}\,\Theta(r-\epsilon)
\nonumber \\
\fl &\quad={\rm w}\!\!\lim_{\epsilon\to 0}\frac{3x_ix_j-r^2\delta_{ij}}{r^5}\Theta(\gamma^2 x_1^2+x_2^2+x_3^2-\epsilon^2)-2\pi[g_{ij}(v)-\case{2}{3}\,\delta_{ij}]\delta(\bi{ r})
\label{identity1}
\end{eqnarray}
and thus the generalized derivative (\ref{genTSr}) can be re-written as
\begin{equation}
\fl \frac{\bar{\partial}^2}{\partial x_i\partial x_j}
\,\frac{1}{r}= {\rm w}\!\!\lim_{\epsilon\to
0}\frac{3x_ix_j-r^2\delta_{ij}}{r^5}\, \Theta(\gamma^2x_1^2+x_2^2+x_3^2-\epsilon^2)-2\pi g_{ij}(v)\delta(\bi{ r}),
\label{genspheroid}
\end{equation}
which is a spheroidal-regularization alternative to the standard, spherical-regularization expression (\ref{genTSr}) for the generalized derivative
$(\bar{\partial}^2/\partial x_i\partial x_j)(1/r)$; (\ref{genspheroid}) reduces to
(\ref{genTSr}) in the limit $v\to 0$ since $\lim_{v\to 0}\gamma=1$ and $\lim_{v\to 0}g_{ij}(v)=\frac{2}{3}\,\delta_{ij}$. An electrodynamic example in which expression (\ref{genspheroid}) is useful is given in Appendix.

\subsection{Cylindrical regularization}
In some applications, cylindrical coordinates are natural to the problem
and the requisite integrations are performed most easily in these coordinates.
We can find the expression for $(\bar{\partial}^2/\partial x_i\partial x_j)(1/r)$ that employs a cylindrical regularization  by evaluating the difference
\begin{eqnarray}
\fl &\lim_{\epsilon\to 0}\int_{r>\epsilon}\rmd^3r\,f(\bi{ r})\frac{3x_ix_j-
r^2\delta_{ij}}{r^5}
-\lim_{\epsilon\to 0}\int_{{\cal T}_{\epsilon,\kappa}}\rmd^3r\,f(\bi{ r})\frac{3x_ix_j-r^2\delta_{ij}}{r^5}\nonumber
\\
\fl &\quad=\lim_{\epsilon\to 0}\left(\int_{{\cal V}_{\epsilon,\kappa}^{(1)}}\rmd^3r\,f(\bi{ r})\frac{3x_ix_j-r^2\delta_{ij}}{r^5}-
\int_{{\cal V}_{\epsilon,\kappa}^{(2)}}\rmd^3r\,f(\bi{ r})\frac{3x_ix_j-r^2\delta_{ij}}{r^5}\right).
\label{diff2}
\end{eqnarray}
Here, the integration domain ${\cal T}_{\epsilon,\kappa}$ is defined as
\begin{equation}
{\cal T}_{\epsilon,\kappa}= \{(x_1,x_2,x_3);|x_1|>\kappa\epsilon \cup x_2^2+x_3^2>\epsilon^2\},
\end{equation}
which is the complement of a cylinder of base radius $\epsilon$ and half-height $\kappa\epsilon$,
parallel to the $x_1$ axis and centered at the origin $\bi{ r}=0$, and the integration domains ${\cal V}_{\epsilon,\kappa}^{(1)}$ and ${\cal V}_{\epsilon,\kappa}^{(2)}$ are defined as
\begin{eqnarray}
\fl {\cal V}_{\epsilon,\kappa}^{(1)}&= \{(x_1,x_2,x_3);x_1^2+x_2^2+x_3^2>\epsilon^2\cap x_2^2+x_3^2<\epsilon^2\cap|x_1|<\kappa\epsilon \}\\
\fl{\cal V}_{\epsilon,\kappa}^{(2)}&= \{(x_1,x_2,x_3);x_1^2+x_2^2+x_3^2<\epsilon^2\cap x_2^2+x_3^2<\epsilon^2\cap|x_1|>\kappa\epsilon \},
\end{eqnarray}
which are regions delimited by the surface of the cylinder and the spherical surface $r^2=x_1^2+x_2^2+x_3^2=\epsilon^2$. The region ${\cal V}_{\epsilon,\kappa}^{(2)}$ is nonempty only when $\kappa<1$, in which case the integral over ${\cal V}_{\epsilon,\kappa}^{(2)}$ has to be subtracted from that over the region ${\cal V}_{\epsilon,\kappa}^{(1)}$.

As $\epsilon$ tends to zero, the regions
${\cal V}_{\epsilon,\kappa}^{(1,2)}$ progressively shrink and collapse onto the origin
$\bi{ r}=0$ so that, for any $\bi{ r}\in {\cal V}_{\epsilon,\kappa}^{(1,2)}$, $f(\bi{ r})\to f(0)$ as $\epsilon\to 0$, and thus
\begin{eqnarray}
\fl &\lim_{\epsilon\to 0}\left[\int_{{\cal V}_{\epsilon,\kappa}^{(1)}}\rmd^3r-\int_{{\cal V}_{\epsilon,\kappa}^{(2)}}\rmd^3r\right]f(\bi{ r})\frac{3x_ix_j-r^2\delta_{ij}}{r^5}\nonumber \\
&\quad =f(0)\lim_{\epsilon\to 0}\left[\int_{{\cal V}_{\epsilon,\kappa}^{(1)}}\rmd^3r-\int_{{\cal V}_{\epsilon,\kappa}^{(2)}}\rmd^3r\right]\frac{3x_ix_j-r^2\delta_{ij}}{r^5},
\end{eqnarray}
where the large brackets are used to denote the difference of the indicated integrals. The integrals on the right-hand side are evaluated easily in cylindrical coordinates
$s, \phi,x_1$. The case $i=j=1$ gives
\begin{eqnarray}
\fl \left[\int_{{\cal V}_{\epsilon,\kappa}^{(1)}}\rmd^3r-\int_{{\cal V}_{\epsilon,\kappa}^{(2)}}\rmd^3r\right]\frac{3x_1^2-r^2}{r^5}
&=4\pi \int_0^{\epsilon}s\,\rmd s\int_{\sqrt{\epsilon^2-s^2}}^{\kappa\epsilon}
\rmd x_1\,\frac{2x_1^2-s^2}{(x_1^2+s^2)^{5/2}}\nonumber\\
\fl &=4\pi\left(\frac{\kappa}{\sqrt{1+\kappa^2}}-\frac{2}{3}\right).
\end{eqnarray}
The cases $i=j=2,3$ give
\begin{eqnarray}
\fl \left[\int_{{\cal V}_{\epsilon,\kappa}^{(1)}}\rmd^3r-\int_{{\cal V}_{\epsilon,\kappa}^{(2)}}\rmd^3r\right]\frac{3x_{2,3}^2{-}r^2}{r^5}
 &=2\int_0^{\epsilon}s\,\rmd s\int_{\sqrt{\epsilon^2{-}s^2}}^{\kappa\epsilon}
\rmd x_1\int_0^{2\pi}\rmd\phi\,\frac{(3\cos^2\phi-1)s^2{-}x_1^2}{(x_1^2+s^2)^{5/2}}
\nonumber\\
\fl &=2\pi\int_0^{\epsilon}s\,\rmd s\int_{\sqrt{\epsilon^2-s^2}}^{\kappa\epsilon}
\rmd x_1\,\frac{s^2-2x_1^2}{(x_1^2+s^2)^{5/2}}\nonumber\\
\fl &=2\pi\left(\frac{2}{3}-\frac{\kappa}{\sqrt{1+\kappa^2}}\right),
\end{eqnarray}
and the mixed cases $i\ne j$ yield zero because of the integration with respect to $\phi$.

Similarly to the establishing of identity (\ref{identity1}), these results
now establish a generalized-function identity
%\begin{equation}
%{\rm w}\!\!\lim_{\epsilon\to 0} %\frac{3x_ix_j-r^2\delta_{ij}}{r^5}[\Theta(r-\epsilon)-\Theta(|x_1|-\kappa\epsilon)
%-\Theta(x_2^2+x_3^2-\epsilon^2)\Theta(\kappa\epsilon-|x_1|)]
%=2\pi[h_{ij}(\kappa)+\tfrac{2}{3}\delta_{ij}]\delta(\bi{ r})
%\end{equation}
\begin{eqnarray}
\fl &{\rm w}\!\!\lim_{\epsilon\to 0}\frac{3x_ix_j-r^2\delta_{ij}}{r^5}\,\Theta(r-\epsilon)\nonumber\\
\fl &\quad={\rm w}\!\!\lim_{\epsilon\to 0}\frac{3x_ix_j-r^2\delta_{ij}}{r^5}\,
[\Theta(|x_1|-\kappa\epsilon)
+\Theta(x_2^2+x_3^2-\epsilon^2)\Theta(\kappa\epsilon-|x_1|)]\nonumber\\
\fl &\quad\quad-2\pi[h_{ij}(\kappa)-\case{2}{3}\delta_{ij}]
\delta(\bi{ r}),
\label{identity2}
\end{eqnarray}
where
\begin{equation}
h_{ij}(\kappa)=\left\{
\begin{array}{ll}
2-2\kappa/\sqrt{1+\kappa^2} &  i=j=1\\
 \kappa/\sqrt{1+\kappa^2}&  i=j=2,3\\
0  &  i\ne j.
\end{array}
\right.
\end{equation}
Using identity (\ref{identity2}), the generalized derivative (\ref{genTSr})
can be written as
\begin{eqnarray}
\fl \frac{\bar{\partial}^2}{\partial x_i\partial x_j}
\,\frac{1}{r}&= {\rm w}\!\!\lim_{\epsilon\to
0}\frac{3x_ix_j-r^2\delta_{ij}}{r^5}\,
[\Theta(|x_1|-\kappa\epsilon)
+\Theta(x_2^2+x_3^2-\epsilon^2)\Theta(\kappa\epsilon-|x_1|)]\nonumber\\
\fl &\quad-2\pi h_{ij}(\kappa)\delta(\bi{ r}).
\label{gencylinder}
\end{eqnarray}
This is a cylindrical-regularization alternative to (\ref{genTSr}). In the limit
$\kappa\to 0$, the delta-function term in (\ref{gencylinder}) simplifies to $-4\pi\delta(\bi{ r})$ for $i=j=1$ (\cite{Feri}, p 29)
and to 0 otherwise; such
regularization is implemented automatically by using the cylindrical
coordinates $s,\phi,x_1$ and performing the requisite integration
over the whole space $\mathbb{R}^3$, but so that the integration
with respect to the variable $x_1$ is done last:
\begin{eqnarray}
\fl \int \rmd^3r\,f(\bi{ r})\frac{\bar{\partial}^2} {\partial x_i\partial x_j}
\,\frac{1}{r}&=\int_{-\infty}^{\infty}\rmd x_1\left(\int_0^{\infty} s\, \rmd s
\int_0^{2\pi}\rmd\phi\,
f(s,\phi,x_1)\frac{d_{ij}(s,\phi,x_1)}{(x_1^2+s^2)^{5/2}}\right)\nonumber\\
\fl &\quad-4\pi f(0)\delta_{i1}\delta_{j1},
\end{eqnarray}
where $d_{ij}(s,\phi,x_1)$ is the function obtained by the transformation of
$3x_ix_j-r^2\delta_{ij}$ from the Cartesian to the cylindrical coordinates
(e.g., $d_{11}(s,\phi,x_1)=2x_1^2-s^2$).

\section{Concluding remarks}
In reference \cite{VH04}, the equivalence of the spherical- and spheroidal-regularization expressions
for the generalized 2nd-order partial derivatives of $1/r$ was illustrated
by a relatively laborious explicit calculation of the weak limit $a\to 0$ of
the derivatives of the Coulomb potential of a charged conducting spheroid of finite extension $a$. The derivation of the generalized derivatives given here, together with the presented results on non-spherical regularization, ensures that any similar explicit calculation must yield the same result.

A point worth making is that while the standard, spherical-regularization   expression (\ref{genTSr})  and the non-spherical-regularization expressions (\ref{genspheroid}) and (\ref{gencylinder})
for the generalized derivatives $(\bar{\partial}^2/\partial x_i\partial x_j)(1/r)$  are guaranteed to yield the same results in an integral with a  well-behaved
function, care should be taken in numerical work to use an integration grid that is compatible with the kind of regularization employed \cite{Feri}.

Very recently,  Frahm's formula (\ref{frahm}) has been criticized as being valid only when averaged over smooth functions, and, to remedy that, an expression
\begin{equation}
\frac{\partial^2}{\partial x_i\partial x_j}\,\frac{1}{r} = \frac{3x_ix_j-r^2\delta_{ij}}{r^5} - 4\pi \frac{x_ix_j}{r^2}\,\delta({\bi r})
\label{franklin}
\end{equation}
has been proposed \cite {Franklin}.
However, our analysis shows that the only `flaw' of Frahm's formula is that it does not indicate explicitly the spherical regularization that it assumes. We note that  expression (\ref{franklin})  still suffers from the lack of an appropriate regularization of the non-delta function term. Moreover, the delta-function term as it stands there is ill-defined; it would become meaningful in an integration in spherical coordinates and the replacement of the 3D delta function  $\delta({\bi r})$
by the radial equivalent $\delta(r)/(4\pi r^2)$. Clearly, the `general' expression (\ref{franklin}) will yield correct results only in an integration in spherical coordinates, with the angular integration of the term involving the non-delta-function part done first.

\ack
The author thanks F Farassat of NASA Langley Research Center for useful discussions on generalized derivatives and helpful comments on a draft.
This paper is written by the author in his
private capacity. No official support or endorsement by the Centers for
Disease Control and Prevention is intended or should be inferred.

\appendix
\section*{Appendix}
\setcounter{section}{1}
We shall employ here a spheroidal-regularization generalized derivative in a calculation of the
difference $A^{({\rm C})}_{x_1}-A^{({\rm L})}_{x_1}$ between the $x_1$-components of the Coulomb- and Lorenz-gauge vector potentials of a unit point charge moving uniformly with a velocity $v$ along the $x_1$ axis.

The difference $A^{({\rm C})}_{x_1}-A^{({\rm L})}_{x_1}$ satisfies an inhomogeneous wave equation \cite{VH_EJP}:
\begin{equation}
\opensquare[A^{({\rm C})}_{x_1}(\bi{ r},t)-A^{({\rm L})}_{x_1}(\bi{ r},t)]
=-v\frac{\partial^2}{\partial x_1^2}\,\phi^{({\rm C})}(\bi{ r},t),
\label{weq}
\end{equation}
where $\opensquare=\nabla^2-\partial^2/\partial t^2$ is the d'Alembertian operator
(we use Gaussian units with the speed of light $c=1$) and
\begin{equation}
\phi^{({\rm C})}(\bi{ r},t)=\frac{1}{\sqrt{(x_1-vt)^2+x_2^2+x_3^2}}
\label{phiC}
\end{equation}
is the Coulomb-gauge scalar potential of the charge.
But an inhomogeneous wave equation
\begin{equation}
\opensquare f=s(x_1-vt,x_2,x_3),
\end{equation}
whose source term is `moving' with a constant velocity $v$ along the $x_1$ axis, can be simplified by a simple transformation of the variables to a Poisson equation, the vanishing-at-infinity solution of which is given in terms of the original variables by
\begin{equation}
\fl f(\bi{ r},t)=-\frac{\gamma}{4\pi}\int \rmd^3{r'}\,\frac{s(x'_1-vt,x'_2,x'_3)}
{\sqrt{\gamma^2(x_1-x'_1)^2+(x_2-x'_2)^2+(x_3-x'_3)^2}},
\end{equation}
where $\gamma=(1-v^2)^{-1/2}$ \cite{PP} (see also \cite{VH_EJP,ODJ}).
The wave equation (\ref{weq}) is therefore solved by
\begin{equation}
\fl A^{({\rm C})}_{x_1}(\bi{ r},t)-A^{({\rm L})}_{x_1}(\bi{ r},t)=
\frac{v\gamma}{4\pi}\int \rmd^3r'\,
\frac{(\bar{\partial}^2/\partial {x'_1}^2)\,\phi^{({\rm C})}(\bi{ r}',t)}
%\frac{\bar{\partial}^2\phi^{({\rm C})}(\bi{ r}',t)/\partial {x'_1}^2}
{\sqrt{\gamma^2(x_1-x'_1)^2+(x_2-x'_2)^2+(x_3-x'_3)^2}}.
\label{AC-AL}
\end{equation}
Note that, in order that the integral is defined properly,  the 2nd-order  derivative of the point-charge potential $\phi^{({\rm C})}$  must be here a generalized one.
However, instead of using the standard spherical-regularization
expression for the derivative, it will be seen that the evaluation of the integral on the right-hand  side of (\ref{AC-AL})
is facilitated greatly when a spherodial-regularization
expression is used.
Following (\ref{genspheroid}), this is given  by
\begin{eqnarray}
\fl \frac{\bar{\partial}^2\phi^{({\rm C})}(\bi{ r},t)}{\partial x_1^2}&=
{\rm w}\!\!\lim_{\epsilon\to 0}
\frac{3(x_1-vt)^2-R^2}{R^5}\,\Theta(R^*-\epsilon)\nonumber\\
\fl &\quad -4\pi\left(\frac{1}{v^2}
-\frac{\arcsin v}{\gamma v^3}\right)\,
\delta(x_1- vt)\delta(x_2)\delta(x_3),
\label{dervphiC}
\end{eqnarray}
where
\begin{equation}
\fl R=\sqrt{(x_1-vt)^2+x_2^2+x_3^2}\quad\mbox{and}\quad R^*=\sqrt{\gamma^2(x_1-vt)^2+x_2^2+x_3^2}.
\label{RR*}
\end{equation}

The integration in (\ref{AC-AL}) is now done in two steps. First, the term involving the delta-function term of ({\ref{dervphiC}) is integrated  readily and  a transformation $\gamma (x'_1-vt) \to x'_1$  is performed in the remaining integrand.  This gives
\begin{equation}
A^{({\rm C})}_{x_1}(\bi{ r},t)-A^{({\rm L})}_{x_1}(\bi{ r},t)=G(\bi{ r},t)
-\left(\frac{\gamma}{v}-\frac{\arcsin v}{v^2}\right)\frac{1}{R^*},
\label{AC-AL_2}
\end{equation}
where
\begin{equation}
\fl G(\bi{ r},t)=
\lim_{\epsilon\to 0}\frac{v}{4\pi}\int \frac{\rmd^3r'}{|\bi{ R}^*-\bi{ r}'|}\,
\frac{2{x'_1}^2/\gamma^2-{x'_2}^2-{x'_3}^2}{({x'_1}^2/\gamma^2+{x'_2}^2+{x'_3}^2)^{5/2}}
\,\Theta(r'-\epsilon).
\label{G}
\end{equation}
Here,
$\bi{ R}^*$ is a vector with components $\gamma (x_1-vt),x_2,x_3$.
 The expansion  of the factor $1/|\bi{ R}^*-\bi{ r}'|$ in  Legendre polynomials now can be used to separate angular and radial  integrations; moreover, the use of spherical coordinates $r', \theta',\phi'$ with the angular integration done first implements the regularization limit $\epsilon\to 0$ automatically. Thus
\begin{equation}
\fl G(\bi{ r},t)
=\frac{v}{2R^*}\sum_{l=1}^{\infty}\frac{(4l+1)P_{2l}(\xi^*)}{2l(2l+1)}
\int_{-1}^1 \rmd\xi'\frac{(3-2v^2){\xi'}^2-1}{(1-v^2{\xi'}^2)^{5/2}}\,P_{2l}(\xi')
\label{G_2},
\end{equation}
where $\xi^*= \gamma(x_1-vt)/R^*$, $\xi'=x'_1/r'= \cos\theta'$, and  the result
\begin{equation}
\int_0^\infty \frac{dr'}{r'}\frac{r_<^{2l}}{r_>^{2l+1}}=
\frac{4l+1}{2l(2l+1)}\,\frac{1}{R^*},\qquad l\ge 1,
\end{equation}
where $r_<$ ($r_>$) is the lesser (greater) of $r'$ and $R^*$, is used.
The summation  in (\ref{G_2}) runs  only over Legendre polynomials of even non-zero order since the integration with respect to $\xi'$  yields zero when the Legendre-polynomial order is zero or odd.

Second, the  $v$-dependent part of the integrand  is expanded  in powers of $v^2$
to facilitate the  integration  with respect to $\xi'$,
\begin{equation}
\fl \frac{(3-2v^2){\xi'}^2-1}{(1-v^2{\xi'}^2)^{5/2}}
=\sum_{n=0}^{\infty}\frac{(2n+1)!!}{(2n)!!}\,
[(2n+3){\xi'}^2-2n-1]{\xi'}^{2n}v^{2n}.
\end{equation}
Using this expansion in (\ref{G_2}), we obtain after interchanging the orders of
summation and integrating term by term with respect to $\xi'$:
\begin{eqnarray}
\fl G(\bi{ r},t)
&=\frac{v}{2\sqrt{\pi}\,R^*}
\sum_{n=0}^{\infty}\,[\Gamma(n+\case{3}{2})]^2v^{2n}\sum_{l=1}^{n+1}
\frac{(4l+1)P_{2l}(\xi^*)}{\Gamma(n-l+2)\Gamma(n+l+\frac{5}{2})}\nonumber\\
\fl &=\frac{1}{vR^*}\sum_{n=1}^{\infty}\frac{[(2n-1)!!]^2}{(2n+1)!}\,
[(2n+1)(v\xi^*)^{2n}-v^{2n}]\nonumber \\
\fl &=\left(\frac{1}{\sqrt{1-v^2\xi^{*2}}}-\frac{\arcsin v}{v}\right)
\frac{1}{vR^*}.
\label{G_3}
\end{eqnarray}
Here, in the 1st line, the series over $l$ terminates at $l=n+1$ since all its $l>n+1$ terms vanish; in the 2nd line, the terminated series is summed and the resulting series over $n$ re-arranged so that it has an overall multiplier $1/v$; and, in the 3rd line, the series over $n$ is summed using the well-known expansions of the functions $(1+x)^{-1/2}$ and $\arcsin x$ in powers of $x$. Using (\ref{G_3}) in (\ref{AC-AL_2}) and then the definition $\xi^*=\gamma(x_1-vt)/R^*$ with the definition
(\ref{RR*}) of $R^*$, we obtain finally
\begin{eqnarray}
\fl &A^{({\rm C})}_{x_1}(\bi{ r},t)-A^{({\rm L})}_{x_1}(\bi{ r},t)
%\fl &\quad=\left(\frac{1}{\sqrt{1-v^2\xi^{*2}}}-\frac{\arcsin v}{v}\right)
%\frac{1}{vR^*}
%-\left(\frac{\gamma}{v}-\frac{\arcsin v}{v^2}\right)\frac{1}{R^*}\nonumber \\
=\left(\frac{1}{\sqrt{1-v^2\xi^{*2}}}-\gamma\right)\frac{1}{v R^*}\nonumber \\
\fl &\quad\quad=\frac{1}{v}\left(\frac{1}{\sqrt{(x_1-vt)^2+x_2^2+x_3^2}}-\frac{1}
{\sqrt{(x_1-vt)^2+(x_2^2+x_3^2)/\gamma^2}}\right).
\end{eqnarray}
The same closed-form expression for the difference between the $x_1$ components of the Coulomb- and Lorenz-gauge vector potentials of a point charge moving uniformly along the $x_1$ axis was obtained in \cite{VH_EJP} by calculating
the requisite gauge function for  the transformation between the Lorenz and Coulomb gauges using a formula derived by Jackson \cite{Jack_AJP}.

%\section*{References}
%\begin{thebibliography}{99}

\end{document}